\documentclass{article}
\usepackage[pdftex]{graphicx}
\usepackage{authblk}
\usepackage{float}
\begin{document}
\date{}
\title{\textbf{2-Loop $\beta$ Function for Non-Hermitian PT Symmetric $\iota g\phi^3$ Theory}}
\author{Aditya Dwivedi\textsuperscript{1}}
\author{Bhabani Prasad Mandal\textsuperscript{2}}
\footnotetext[1]{ email address: adityadwivedi0224@gmail.com}
\footnotetext[2]{ email address: bhabani.mandal@gmail.com, bhabani@bhu.ac.in
 }
\affil{\small{Department of Physics, Institute of Science, Banaras Hindu University}}
\affil{\small{Varanasi-221005}}
\affil{\small{INDIA}}
      \maketitle
    \begin{abstract}
  We investigate Non-Hermitian quantum field theoretic model with $\iota g\phi^3$ interaction in 6 dimension. Such a model is PT-symmetric for the pseudo scalar field $\phi$. We analytically calculate the 2-loop $\beta$ function and analyse the system using renormalization group technique. Behavior of the system is studied near the different fixed points. Unlike $g\phi^3$ theory in 6 dimension $\iota g\phi^3$ theory develops a new non trivial fixed point which  is energetically stable.  
    \end{abstract}
\section{Introduction}
Over the past two decades a new field with combined Parity(P)-Time reversal(T) symmetric non-Hermitian systems has emerged and has been one of the most exciting topics in frontier research. It has been shown that such theories can  lead to the consistent quantum theories with real spectrum, unitary time evolution and probabilistic interpretation in a different Hilbert space equipped with a  positive definite inner product \cite{real1}-\cite{real3}. The huge success of such non-Hermitian systems has lead to extension to many other branches of physics and interdisciplinary areas. The novel idea of such theories have been applied in numerous systems leading huge number of application \cite{pt4}-\cite{shal2}. 
\par       Several PT symmetric non-Hermitian models in quantum field theory have also been studied in various context \cite{pt}-\cite{qft2}. Deconfinment to confinment transition is realised by PT phase transition in QCD model using natural  but unconventional hermitian property of the ghost fields \cite{pt}. PT symmetric quantum field theory involving non-Hermitian mass term $\alpha\bar{\psi}\gamma^5\psi$ was introduced and further investigated to show the existence of conserved current in such a model to show the consistency of PT symmetry and unitarity of the theory. Theories with non-Hermitian mass term has also been used for alternative description of neutrino mass and dark matter. Neutrino oscillation has been investigated using non-Hermitian PT symmetric model in quantum field theory. Aspects of spontaneous symmetry breaking and Goldston theorem has been studied using non-Hermitian field theoretic model \cite{x}.
\par Pseudo scalar field theories with $\iota g\phi^3$ interaction has been investigated by several groups \cite{sha0}, \cite{sha00}. The motivation behind the study of $\iota g\phi^3$ is stated as follows. In ordinary $g\phi^3$ theory ground state is unstable and it decays gradually, whereas in $\iota g\phi^3$ potential is complex.  Therefore concept of boundedness of ground state is not going to apply here as explained in Refs.\cite{re1}-\cite{re3}.   One can calculate the ground state energy density  for $\iota g\phi^3$ theory by using perturbation theory and summing all the connected vacuum Feynman diagrams. The Borel summation of  this Stieltjes series is real \cite{re1}-\cite{re3}. Thus, in view of this one can say that ground state of this theory is stable and $\iota g\phi^3$ theory is physically acceptable.
Further in $1+1$ dimension $\iota g\phi^3$ theory provides  an important feature in their solutions. The field $\phi$ is pure imaginary soliton wave, which can be useful in the description of hadrons, because they does not change shape after interaction \cite{sol1}. 
Due to boundary condition  $\phi_{\pm\infty}=0$ and $\phi_{\pm\infty}^\dagger=0$, we can say that these are non topological solitons and these characterise the theory with more than one component. Merging of $\iota g\phi^3$ and $-\iota g\phi^3$ theory is equivalent to Lee-Wick theory which includes both $\phi$ and $\phi^\dagger$ \cite{sol2}.
Such a theory is shown to Hermitian equivalent to Lee-Wick theory which suffers from the existence of the famous ghost state and instability problem \cite{shal}. Metric operator has been computed for $\iota\phi^3$ theory by making new ansatz \cite{shal2}. The critical behavior of such theories around the fixed points has been investigated using renormalization group technique at the one loop order \cite{bend1}. Further renormalization group properties of ordinary $\phi^3$ theory have compares with that of PT symmetric $\iota g\phi^3$ theory in 1-loop $\beta$ function for the later theory  \cite{bend2}.
\par The purpose of the present article is to revisit the $\iota g\phi^3$ theory in 6 dimension in the framework of renormalization group analysis and to examine the theory with higher loop calculation. We calculate  2-loop $\beta$ function explicitly and analyse the behavior of the system near the fixed points. Unlike the usual $ g\phi^3$  theory the PT symmetric $\iota g\phi^3$ theory develops non-trivial fixed point which is energetically stable. The theory is perturbatively renormalizable. Our analysis is  consistent with renormalization group study of the same model at the 1-loop level \cite{bend2}.\\
\par Now, we present the plan of our paper. In Section 2 we discuss $g\phi^3$ theory in the framework of RG analysis. earlier. 1-loop calculations of $\beta$ function is carried out in Section 3. 2- loop calculation are presented in Section 4. Section 5 is kept for results and discussions.
\section{Theory of $\phi^3$ fields:} 
We start with the bare Lagrangian density of  $\phi^3$ theory in 6 dimension as:
\begin{equation}
\L=\frac{1}{2}(\partial\phi_0)^2-\frac{1}{2}{m_0^2\phi_0^2}-\frac{1}{6}{g_0\phi_0^3}
\end{equation} 
where the subscript 0 denotes the bare quantities .
This Lagrangian density further can be expressed in terms of renormalized Lagrangian density as 
$$
L=\frac{1}{2}(\partial\phi)^2-\frac{1}{2}m^2\phi^2-\frac{1}{6}{g}{\phi^3}+\frac{1}{2}{C}(\partial\phi)^2-\frac{1}{2}B\phi^2-\frac{1}{6}A\phi^3. 
$$
Where m, g and $\phi$ are renormalized mass, coupling constant and field respectively and satisfy:
$$\sqrt{Z}\phi=\phi_0$$
$${g_0}={Z_0}{Z^{-\frac{3}{2}}}{g}.$$
The coefficient of counter Lagrangian are written as $Z=1+C$, ${m_0^2}Z=m^2+B$, ${Z_0}g=g+A$.\\
In the arbitrary n dimension the coupling $g$ is dimensionful, to make it dimensionless a mass scale $\mu$ is introduced.
Unrenormalized coupling and mass are divergent in $6$ dimension therefore these quantities can be expanded around $n=6$ in  terms of renormalized parameter as
\begin{equation}
g_0Z^\frac{3}{2}=\mu^\frac{6-n}{2}\left[g+\sum_{\alpha=1}^{\infty}\frac{e_\alpha(m,g)}{(n-6)^\alpha}\right] 
\end{equation}
\begin{equation}
m_0^2Z=\mu\left[m^2+\sum_{\alpha=1}^\infty\frac{f_\alpha(m,g)}{(n-6)^\alpha}\right]
\end{equation} and
\begin{equation}
Z=1+\sum_{\alpha=1}^\infty\frac {h_\alpha(m,g)}{(n-6)^\alpha}
\end{equation}
where $e_\alpha$, $f_\alpha$ and $h_\alpha$ are the different  coefficients in the above expansion for coupling, mass and Z terms for the pole at $n=6$ of order  $\alpha$.
To check the uniqueness of the above series let us write $g$ and $m$ in form of\\
$$
g\to\left[g+x_1(n-6)+x_2(n-6)^2+......\right]$$
and 
$$m\to\left[m+y_1(n-6)+y_2(n-6)^2+......\right].$$
On putting these in equations (2) and (3) we obtain 
\begin{equation}
g_0Z^\frac{3}{2}=\mu^\frac{6-n}{2}\left[g+\sum_{k=1}^\infty x_k(n-6)^k+\sum_{\alpha=1}^\infty\frac{e_\alpha(m',g')}{(n-6)^\alpha}\right]
\end{equation}
\begin{equation}
m_0^2Z=\mu\left[m^2+\sum_{k=1}^\infty y_k(n-6)^k+\sum_{\alpha=1}^\infty\frac{f_\alpha(m'g')}{(n-6)^\alpha}\right]
\end{equation}
\begin{equation}
Z=1+\sum_{\alpha=1}^\infty\frac {h_\alpha(m',g')}{(n-6)^\alpha}
\end{equation}
These equations are different from those which we first considered in equation (2) and (3). For the uniqueness of the series we can put the additional condition that the coefficients of $x_k$ and $y_k$ are 0 in equations (5) and (6). In other words to say  that all the transformations of $g$ and $m$ always lead to the series in equations (2),(3) and (4).
Since mass parameter $\mu$  is arbitrary we have to check the uniqueness of the series in equations (2) and (3) after changing it too. In order to do that we shift $\mu$ as $\mu'=\mu(1+\epsilon)$ ,where $\epsilon<<1$.
Putting $\mu'$ in equations (2) and (3) we obtain -
$$g_0Z^\frac{3}{2}=\mu'^{(\frac{6-n}{2})}\left[g+g(\frac{n-6}{2})\frac{\epsilon}{2}+\frac{\epsilon}{2}e_1+\sum_{\alpha=1}^{\infty}\frac{e_\alpha+\frac{\epsilon}{2}e_{\alpha+1}}{(n-6)^\alpha}\right]$$ and
$$m_0^2Z=\mu'(1-\epsilon)\left[m^2+\sum_{\alpha=1}^\infty\frac{f_\alpha}{(n-6)^\alpha}\right ].$$
These equations will be put back in the form of equations (2) and (3) when  -
$$g=\tilde{g}-\frac{\epsilon}{2}(n-6).$$ 
This transformation makes the above series as
\begin{equation}
g_0Z^\frac{3}{2}=(\mu')^\frac{6-n}{2}\left[\tilde{g}+\frac{\epsilon}{2}e_1-\frac{\epsilon}{2}\tilde{g}e_{1,g}+\sum_{\alpha=1}^\infty\frac{e_\alpha(m,\tilde{g})+\frac{\epsilon}{2}e_{\alpha+1}-\frac{\epsilon}{2}\tilde{g}e_{\alpha+1,g}}{(n-6)^\alpha}\right  ]
\end{equation}
\begin{equation}m_0^2Z=\mu'\left[m^2-\epsilon m^2-\frac{\epsilon}{2} \tilde{g} f_{1,g}+\sum_{\alpha=1}^\infty  \frac{f_\alpha(m,\tilde{g})-\frac{\epsilon}{2}f_\alpha-\frac{\epsilon}{2}f_{\alpha+1,g}}{(n-6)^\alpha}\right ] 
\end{equation}
where $e_{\alpha,g}$ $\equiv$ $\frac{\partial{e_\alpha}}{\partial\tilde{g}}$ and similarly for others.
We see that changing the mass scale, we have to change the renormalized parameters also. As at n=6, $\tilde{g}=g$, thus new renormalized parameter in terms of old parameters in view of above equations  can be written as
\begin{equation}
g'=g+\frac{\epsilon}{2}\left[1-g\frac{\partial}{\partial g}\right]e_1(m,g)
\end{equation} and
\begin{equation}
m'^2=m^2-\epsilon \left[m^2+\frac{g}{2}\frac{\partial}{\partial g}f_1(m,g)\right]
\end{equation}
It  has been shown that $e_\alpha$, $f_\alpha$ and $h_\alpha$ are independent of mass, thus above equation can be written as-
\begin{equation}
g'=g+\frac{\epsilon}{2}\beta_g
\end{equation} 
\begin{equation}
m'^2=m^2-\epsilon \gamma_{m^2}
\end{equation}
where $\gamma_{m^2}=\left[m^2+\frac{g}{2}\frac{\partial}{\partial g}f_1(m,g)\right]$ and, $\beta$ function for the coupling is  defined as
\begin{equation}
\beta_g=\big[1-g\frac{\partial}{\partial g}]e_1.
\end{equation}
On deriving the renormalization group equation for n point green function we get the result \begin{equation}
\mu\frac{\partial{m^2}}{\partial\mu}=\beta_{m^2}=-m^2\big[1+\frac{g}{2}\frac{\partial}{\partial g}\big]f_1.
\end{equation}
From equations (14) and (15) we can  see that if we use the 't Hooft procedure \cite{thooft1}, \cite{thooft2}, the beta function depends only on the one loop pole coefficients. Beta function upto higher order in coupling is found by relating  higher order pole coefficients to the single pole coefficients. This shows that higher order pole coefficients are not arbitrary and depends on the one loop pole coefficients. It is well known that if theory is renormalizable in one loop order then it is renormalizable at higher loops also.
In view of the above, the series in equation (2), (3) and (4) are changed to the form
\begin{equation}
g_0Z^\frac{3}{2}=\mu^\frac{6-n}{2}\left[g+\sum_{\alpha=1}^\infty \frac{e_\alpha(g)}{(n-6)^\alpha}\right]=\mu^\frac{6-n}{2}\left[g+g\sum_{\alpha=1}^\infty \sum_{\gamma=\alpha}^\infty \frac{e_{\alpha\gamma} g^\gamma }{(n-6)^\gamma}\right]
\end{equation}
\begin{equation}
m_0^2Z=\mu\left[m^2+m^2\sum_{\alpha=1}^\infty \frac{f_\alpha(g)}{(n-6)^\alpha}\right]=\mu\left[m^2+m^2\sum_{\alpha=1}^\infty \sum_{\gamma=\alpha}^\infty \frac{f_{\alpha\gamma} g^\gamma}{(n-6)^\alpha}\right]
\end{equation}
and\begin{equation}
Z=1+\sum_{\alpha=1}^\infty \frac{h_\alpha(g)}{(n-6)^\alpha}=1+\sum_{\alpha=1}^\infty \sum_{\gamma=\alpha}^\infty \frac{h_{\alpha\gamma}g^\gamma }{(n-6)^\alpha}.
\end{equation}
Further, by considering one and two loop diagrams, we will calculate the terms in these series expansion taking $\alpha=1$ and 2 and $\gamma=2,4$ which will enable us to calculate the coefficients of the pole terms. This will in term help us to find the $\beta$ function for the theory up to two loops.\\
\section{One loop calculation:}
We want to use the general formula written in the previous section to calculate the amplitudes explicitly for different graphs in $g\phi^3$ theory in one loop order. Let us consider the following diagrams and counter terms relevant to the calculation of $\beta$ function in one loop order.\\
\begin{figure}[h!]
\centering
\includegraphics[scale=.4]{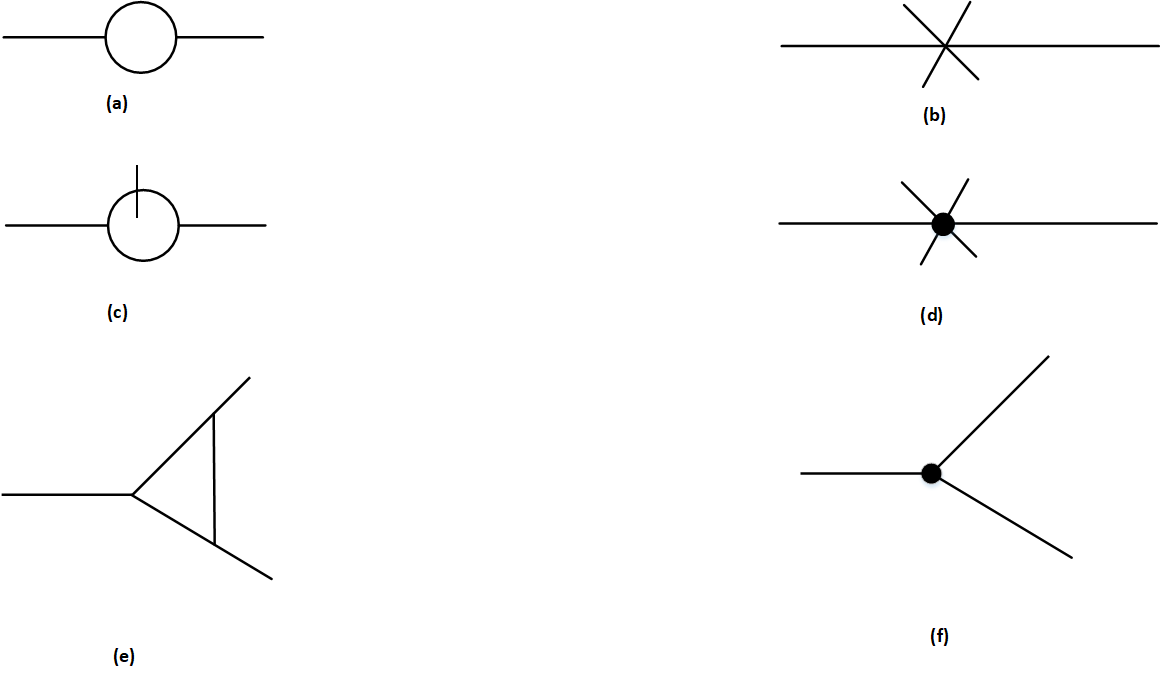}
\caption{(a) One loop self energy diagram, (b) Wave function counter term, (c) Self energy with massive propagator, (d) Mass counter term, (e) Diagram for vertex term, (f) Vertex counter term}
\end{figure}\\
We calculate the amplitudes for these diagrams using the technique given in reference  \cite{woo}.
  Diagram $1(a)$ is self energy diagram  
which  gives the result \\
\begin{equation}
A_{1(a)}=\frac{g^2p^2}{12(4\pi)^3}\left[\frac{2}{(n-6)}-\ln\frac{-p^2}{4\pi\mu^2}-(\gamma-\frac{8}{3})\right].
\end{equation}
Since the  amplitudes of massless two point diagrams are related to the field renormalization coefficient Z. We use equation (18), which suggest that term corresponding to diagram 1(b) is $\frac{1}{2}\left[\frac{h_{12}}{n-6}\right]g^2(\partial\phi)^2$, and amplitude for the diagram is $\left[\frac{h_{12}}{n-6}\right]g^2p^2$. On comparing the pole part of $A_{1(a)}$ with amplitude of diagram $1(b)$ we get, $ h_{12}=\frac{1}{6(4\pi)^3}$. The contribution of 1(c) diagram is \\
\begin{equation}
A_{1(c)}=\frac{m^2 g^2}{(4\pi)^\frac{n}{2}}\Gamma(3-\frac{n}{2})\big(\frac{-p^2}{\mu^2}\big) B(\frac{n}{2}-1,\frac{n}{2}-1).
\end{equation}
Pole part in $A_{1(c)}$ is present due to the pole of gamma function present in the amplitude.
Amplitudes of massive two point diagrams are related to the mass renormalization. Using equation (17), we get contribution of diagram 1(d) in lagrangian density is $-\frac{1}{2}f_{12}\left[\frac{m^2g^2\phi^2}{(n-6)}\right]$ and we compare the pole part of $A_{1(c)}$ to the amplitude of $1(d)$ to get $f_{12}=\frac{1}{(4\pi)^3}$. Amplitude of the diagram $1(e)$ is $$A_{1(e)}=\frac{g^3\mu^\frac{\epsilon}{2}}{2(4\pi)^3}\frac{2}{n-6}$$. Three point diagrams are related to the renormalization of the coupling of the theory thus, use of equation (16) suggest that diagram $1(f)$ corresponds to the term $-\frac{1}{6}\left[\frac{e_{12}}{n-6}\right]g^2\phi^2$ and on comparing the  coefficient of simple pole of $A_{1{e}}$ with amplitude of diagram  $1(f)$ gives $e_{12}=\frac{1}{(4\pi)^3}$.
\section{Two Loop calculation}
In this section we consider all the two loop diagrams in the $g\phi^3$  theory in 6 dimension  which are relevant to our calculation for $\beta$ function. First we consider 2 loop self energy diagrams (Figs. 2(a), 2(b))  and self energy counter terms (Figs. 2(c), 2(d)). Then we consider the diagrams with massive propagator and, finally we consider 2 loop three point diagrams for finding the various pole coefficients.
\subsection{Massless self energy diagrams }
 We have to consider following diagrams to calculate the pole coefficients of two loop level for the self energy diagram for massless case.
\begin{figure}[h!]
\centering
\includegraphics[scale=.4]{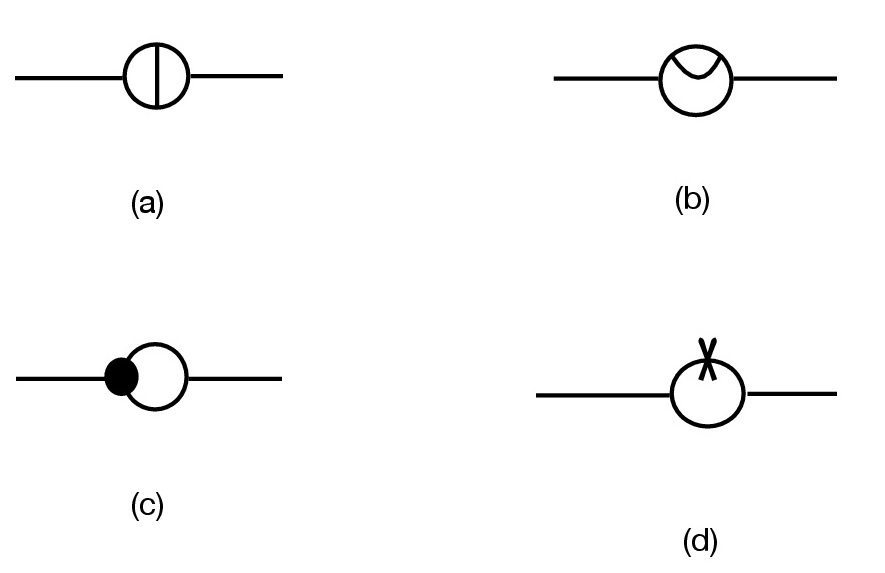}
\caption{Two loop diagram for massless case. (a) Self energy with overlapping divergences, (b) Self energy with nested divergences, (c) Self energy with vertex counter term , (d) Self energy with wave function counter term.}
\end{figure}\\

On analysis of these, we can easily see that there are overlapping divergences in these diagrams. After doing the separation of overlapping divergences, we get the contribution from various diagrams.
Amplitude for the diagram 2(a) is evaluated as
$$A_{2(a)}=\frac{g^4p^2}{4\pi^6}\left[-\frac{1}{6(n-6)^2}-\frac{\ln(\frac{-p^2}{\mu^2})}{6(n-6)}+\frac{1}{6}\frac{(-\gamma+3)}{n-6}\right]+ \mbox {finite part}.$$\\
Similarly we find the amplitudes for diagrams 2(b), 2(c) and 2(d) which are written as-
$$A_{2(b)}=\frac{2g^4p^2}{(4\pi)^6}\left[\frac{1}{36(n-6)^2}+\frac{ln(\frac{-p^2}{\mu^2})}{36(n-6)}+\frac{12\gamma-43}{432(n-6)}\right]+\mbox {finite part}$$\\
$$A_{2(c)}=\frac{g^4p^2}{(4\pi)^6}\left[\frac{1}{3(n-6)^2}+\frac{ln(\frac{-p^2}{\mu^2})}{6(n-6)}+\frac{3\gamma-8}{18(n-6)}\right]+\mbox {finite part}$$\\
$$A_{2(d)}=\frac{g^4p^2}{(4\pi)^6}\left[-\frac{1}{18(n-6)^2}-\frac{ln(\frac{-p^2}{\mu^2})}{36(n-6)}-\frac{3\gamma-8}{108(n-6)}\right]+\mbox {finite part}$$\\
On adding these amplitudes corresponding to the diagrams 2(a), 2(b), 2(c) and 2(d), we see that logarithmic divergent parts vanish and on comparing  the pole parts of the added result using  equation (18) and diagrams which give counter terms, as done in the previous section, we get -
\begin{equation}
\frac{h_{14}}{n-6}=\frac{13}{432(4\pi)^6(n-6)}
 \end{equation} 
and 
\begin{equation}\frac{h_{24}}{(n-6)^2}=\frac{5}{36(4\pi)^6(n-6)^2}.
\end{equation}
Next we proceed to evaluate the massive case.
\subsection{Contribution to mass term}
Following diagrams contribute to the mass term pole coefficients in the two loop order.
\begin{figure}[h!]
\includegraphics[scale=.4]{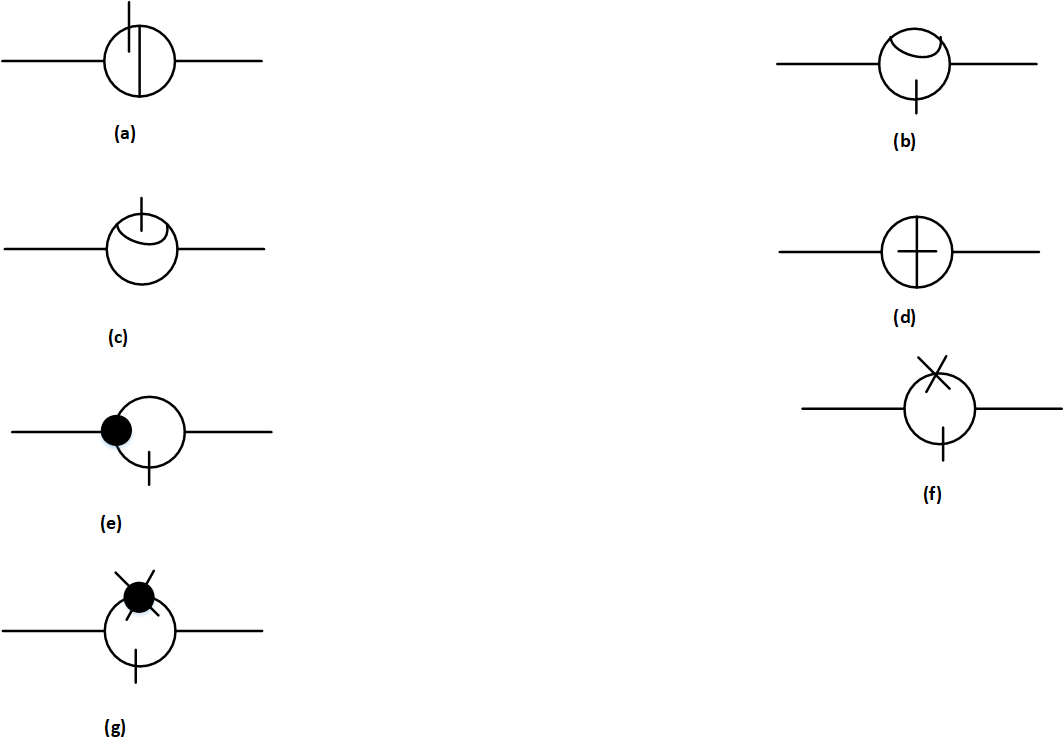}
\caption{Two loop diagrams for massive case. 3(a), 3(b), 3(c) and 3(d) are  self energy diagrams with massive propagator, (e) self energy vertex counter term with massive propagator, (f) self energy   wave function counter term with massive propagator, (g) self energy mass counter term with massive propagator.}
\end{figure}
Contribution for these diagrams are calculated explicitly and given below.\\ Amplitude for the diagram 3(a) is 
$$A_{3(a)}=\frac{g^4m^2}{(4\pi)^6}\left[\frac{1}{(n-6)^2}+\frac{ln(\frac{-p^2}{\mu^2})}{(n-6)}+\frac{4\gamma-9}{4(n-6)}\right]+\mbox {finite part}.$$
 Amplitude for the diagrams 3(b), 3(c), 3(d), 3(e), 3(f) and 3(g) are 
$$A_{3(b)}=\frac{g^4m^2}{(4\pi)^6}\left[-\frac{1}{12(n-6)^2}-\frac{ln(\frac{-p^2}{\mu^2})}{12(n-6)}-\frac{12\gamma-31}{144(n-6)}\right]+\mbox {finite part}$$
$$A_{3(c)}= \frac{g^4m^2}{(4\pi)^6}\left[\frac{1}{2(n-6)^2}+\frac{ln(\frac{-p^2}{\mu^2})}{2(n-6)}+\frac{(4\gamma-9)}{8(n-6)}\right]+\mbox {finite part}$$
$$A_{3(d)}=\frac{g^4m^2}{(4\pi)^6}\left[-\frac{1}{4(n-6)}\right]+\mbox {finite part}$$
$$A_{3(e)}=\frac{g^4m^2}{(4\pi)^6}\left[-\frac{2}{(n-6)^2}-\frac{ln(\frac{-p^2}{\mu^2})}{(n-6)}-\frac{\gamma+2}{(n-6)}\right]+\mbox {finite part}$$
$$A_{3(f)}=-\frac{1}{12}\frac{g^4m^2}{(4\pi)^6}\left[-\frac{2}{(n-6)^2}-\frac{ln(\frac{-p^2}{\mu^2})}{(n-6)}-\frac{\gamma+2}{(n-6)}\right]+\mbox {finite part}$$
$$A_{3(g)}=\frac{g^4m^2}{(4\pi)^6}\left[-\frac{1}{(n-6)^2}-\frac{ln(\frac{-p^2}{\mu^2})}{2(n-6)}-\frac{\gamma+2}{2(n-6)}\right]+\mbox {finite part}$$\\ respectively.
Adding up all these contributions we see that the logarithmic part in the amplitudes cancels each other and then comparing the added amplitude using equation (17) we get, 
\begin{equation}
\frac{f_{14}}{(n-6)}=\frac{23}{48(n-6)(4\pi)^6}
\end{equation}
 and 
\begin{equation}
\frac{f_{24}}{(n-6)^2}=\frac{5}{4(n-6)^2(4\pi)^6}.
\end{equation}
\subsection{2 loop three point diagrams}
In this section we discuss the 2 loop three point diagrams which are relevant for calculation of $\beta$ function, these diagrams are given in  figure 4.
\begin{figure}[h!]
\includegraphics[scale=.5]{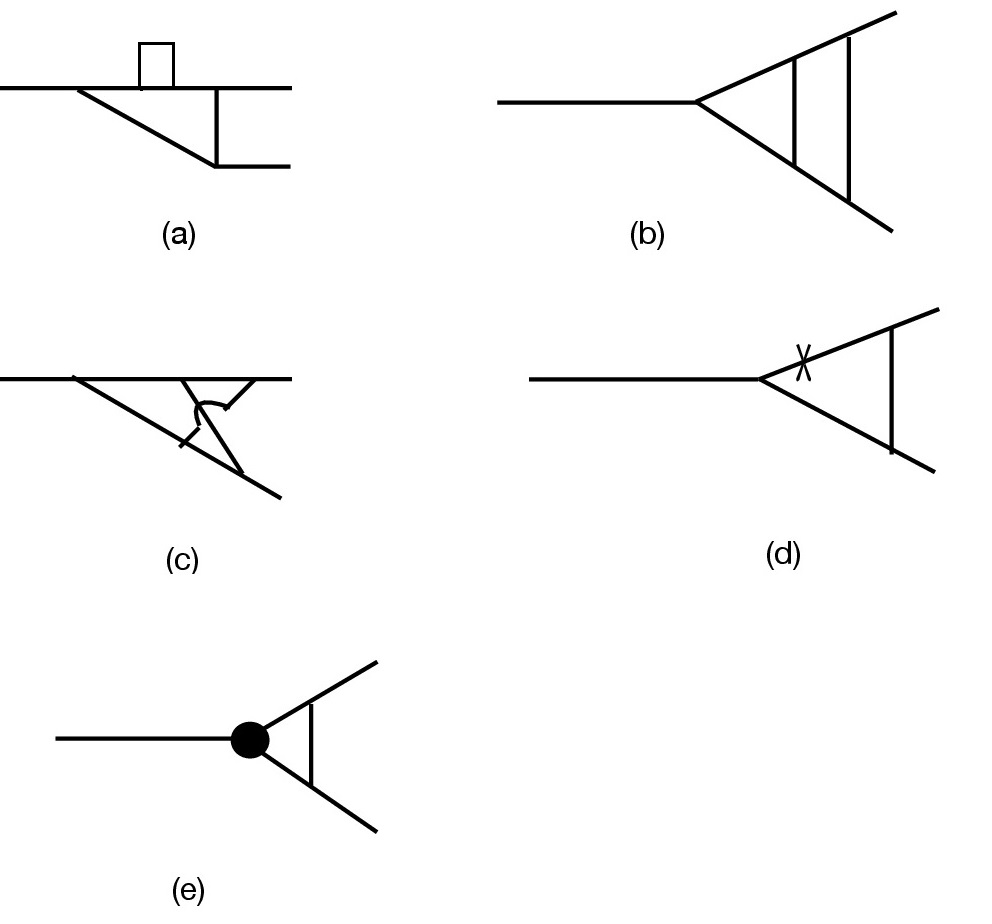}
\caption{Two loop three point diagram 4(a), 4(b) and 4(c) represent vertex term, 4(d)  vertex term with wave function counter term, 4(e) vertex term with vertex counter term.}
\end{figure}
Amplitudes for 2 loop three point diagrams are given as- 

$$A_{4(a)}=\frac{g^4}{(4\pi)^6}\left[-\frac{1}{4(n-6)^2}-\frac{ln(\frac{q^2}{\mu^2})}{4(n-6)}-\frac{12\gamma-7}{48(n-6)}-\frac{X}{2(n-6)}\right]+\mbox {finite part}$$
$$A_{4(b)}=\frac{g^4}{(4\pi)^6}\left[\frac{3}{2(n-6)^2}+\frac{3}{2}\frac{ln(\frac{q^2}{\mu^2})}{(n-6)}-\frac{12\gamma-3}{8(n-6)}+\frac{3X}{(n-6)}\right]+\mbox {finite part}$$
$$A_{4(c)}=-\frac{g^4}{4(4\pi)^6}\frac{1}{n-6}+\mbox {finite part}$$
$$A_{4(d)}=\frac{g^4}{(4\pi)^6}\left[\frac{1}{2(n-6)^2}+\frac{ln(\frac{q^2}{\mu^2})}{4(n-6)}-\frac{\gamma}{4(n-6)}+\frac{X}{2(n-6)}\right]+\mbox {finite part}$$
$$A_{4(e)}=\frac{g^4}{(4\pi)^6}\left[-\frac{3}{(n-6)^2}-\frac{3}{2}\frac{ln(\frac{q^2}{\mu^2})}{(n-6)}-\frac{\gamma}{2(n-6)}-\frac{3X}{n-6}\right]+\mbox {finite part}$$
respectively,
where $X=\int_0^1{\delta{x}\delta{y}\delta{z}\delta(1-x-y-z)\ln(xy+yz+zx)}$.\\
Again adding up all the contribution we see that the logarithmic part is removed and, then on comparing using equation (16) we get the pole term  coefficients
 \begin{equation}
 \frac{e_{14}}{n-6}=\frac{23}{48(4\pi)^6(n-6)}
 \end{equation}
 and 
 \begin{equation}
 \frac{e_{24}}{(n-6)^2}=\frac{5}{4(n-6)^2(4\pi)^6}.
 \end{equation}
\section{Calculation of $\beta$ function for $\iota{g}{\phi^3}$ theory:}
Now we are ready to calculate the 2 loop $\beta$ function for the theory using equation (14) and other results in the previous section.
To calculate the value of $e_1$, we use equations (2) and (16). We expand both series and compare the coefficients of $\frac{1}{n-6}$ which give us the relation-
\begin{equation}
e_1=(e_{12}-\frac{3}{2}h_{12})g^3+(e_{14}-\frac{3}{2}h_{14})g^5.
\end{equation}
Now using equation (14), for $g\phi^3$ theory, we have
\begin{equation}
\beta(g)=-\frac{3}{4}\frac{g^3}{(4\pi)^3}-\frac{125}{144}\frac{g^5}{(4\pi)^6}.
\end{equation}\\
For  $\iota g\phi^3$ theory, we replace $g$ by $\iota g$ in the above expression of the $\beta$ function, which give
\begin{equation}
\beta(\iota{g})=\frac{3}{4}\frac{\iota{g^3}}{(4\pi)^3}-\frac{125}{144}\frac{\iota{g^5}}{(4\pi)^6}.    
\end{equation}\\
The fixed points of theory are obtained by putting $\beta(\iota{g})=0$, and given by
,
\begin{equation}
g=\pm\sqrt{\frac{108(4\pi)^3}{125}}\equiv \tilde{g}(say)\ \mbox{and}, \  g=0.
\end{equation}
Now to see how the system behaves near the fixed point, we revisit equation(14),
 $$\mu\frac{\partial{ g}}{\partial{\mu}}=\beta{(g)}$$
 Integrating this, using $\beta$ function given in equation (29) 
 $$ \int\frac{\partial g}{\alpha{g^3}-\lambda{g^5}}=\int \frac{\partial \mu}{\mu} $$
 where $\alpha\equiv\frac{3}{4}\frac{1}{(4\pi)^3}$ and $\lambda\equiv\frac{125}{144}\frac{1}{(4\pi)^6}$.
 Solution of the above integral is obtained as-
 $$\mu=\exp (\frac{-1}{\alpha}\frac{1}{2g^2})\big[ \frac{g}{(1-\frac{\lambda}{\alpha}{g^2})^\frac{1}{2}}\big]^{ ( \frac{\lambda}{\alpha^2})}.$$\\
This indicates that
$g=\sqrt{\frac{\alpha}{\lambda}}=\sqrt{\frac{108(4\pi)^3}{125}}=\tilde{g}$ is a stable fixed point as $ g\to\tilde{g}$, $\mu$ grows.
In the high energy range, properties of  $\iota{g}\phi^3$ theory in 6 dimension is governed by the RG trajectories near the fixed point $g=\tilde{g}$. We would like to point out that in  $\phi^3$ theory in 6 dimension the only fixed point $g=0$ was the trivial one. 
In case of low energy limits i.e. $\lim{\mu\to 0}$ the coupling ${g\to 0}$ which shows that theory behaves as a free theory in low energy limit and behavior of the theory is governed by the trajectories near the Gaussian fixed point $g=0$.
\begin{figure}[H]
\centering
\includegraphics[scale=.5]{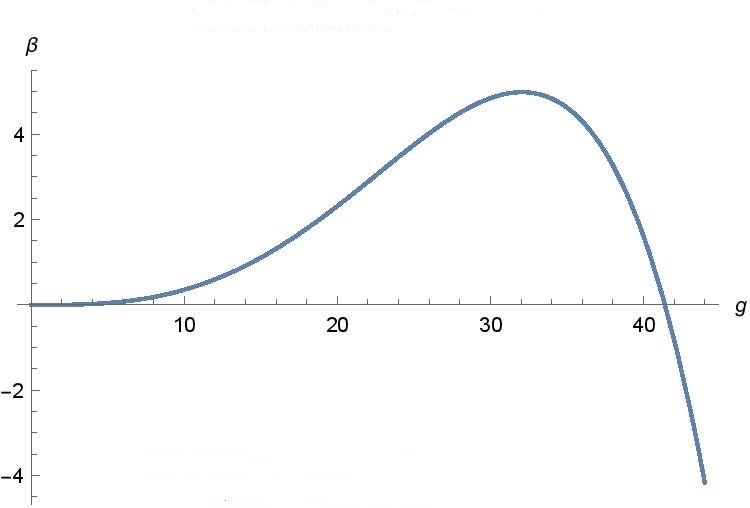}
\caption{Plot of $\beta$ function of coupling $g$ for $\iota g \phi^3$ theory.}
\end{figure}
\begin{figure}
\centering
\includegraphics[scale=.5]{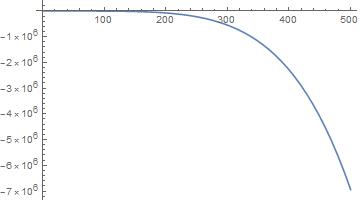}
\caption{Plot of $\beta$ function of coupling $g$ for $g\phi^3$ theory.}
\end{figure}
The variation of $\beta$ function with coupling $g$ for $\iota g \phi^3$ is shown in fig (5).
This behaviour of $\beta$ function is then compared with that of usual $ g \phi^3$ theory, given in fig (6).
We find the dependence of mass on coupling using equation (15). To find the value of $f_1$ we use equation (3) and (17).
We calculate the $\beta_{m^2}$ for the theory with $\iota g\phi^3$ coupling as,
\begin{equation}
\beta_{m^2}=-m^2\big[1-\frac{5}{12}\frac{g^2}{(4\pi^3)}+\frac{97}{216}\frac{g^4}{(4\pi)^6}\big].
\end{equation}
Equation (31) implies that $m^2=0$ is only fixed point for the $\beta_{m^2}.$\\
Using the definition of the beta function,
\begin{equation}
\mu\frac{\partial m^2}{\partial\mu}=-m^2\big[1-\frac{5}{12}\frac{g^2}{(4\pi^3)}+\frac{97}{216}\frac{g^4}{(4\pi)^6}\big]
\end{equation}
we found that
$$\frac{1}{m^2}=\mu\exp\big[1-\frac{5}{12}\frac{g^2}{(4\pi^3)}+\frac{97}{216}\frac{g^4}{(4\pi)^6}\big]$$
which shows that on going $\lim{\mu\to\infty}$ we see that ${m^2\to 0}$ which shows that $m^2=0$ is stable fixed point for the beta function for the mass term.
\section{Conclusion}
In this work, we have investigated a PT symmetric non-Hermitian model for scalar fields in 6 dimension  to extract some new features of the theory. By considering the appropriate Feynman diagrams in one loop and two loop order and the relevant counter terms for those diagrams, we explicitly evaluate two loop $\beta$ finction for $\iota g \phi^3$ theory in $6$ dimension.Unlike the Hermitian theory, this non-Hermitian theory develops additional non-trivial energetically stable fixed point. The high energy behavior of the theory is governed by this non trivial fixed point (fig 5). On the other hand low energy behaviour is governed by the RG trajectories near the Gaussian fixed point $g=0$. $\beta$ function for the mass term for this non-Hermitian model has been calculated. $m^2=0$ is a stable fixed point for the $\beta$ function for mass term.  

\end {document}